# SIMULATION OF THERMAL AND CHEMICAL RELAXATION IN A POST-DISCHARGE AIR CORONA REACTOR

M. Meziane, J.P. Sarrette, O. Eichwald, O. Ducasse and M. Yousfi

University of Toulouse, LAPLACE, UMR CNRS 5213, 118 route de Narbonne, Bât. 3R2, 31062 Toulouse Cedex 9, France- email: eichwald@laplace.univ-tlse.fr
.

## ABSTRACT

In a DC point-to-plane corona discharge reactor, the mono filamentary streamers cross the inter electrode gap with a natural repetition frequency of some tens of kHz. The discharge phase (including the primary and the secondary streamers development) lasts only some hundred of nanoseconds while the post-discharge phases occurring between two successive discharge phases last some tens of microseconds. From the point of view of chemical activation, the discharge phases create radical and excited species located inside the very thin discharge filaments while during the post-discharge phases these radical and excited species induce a chemical kinetics that diffuse in a part of the reactor volume. From the point of view of hydrodynamics activation, the discharge phases induce thermal shock waves and the storage of vibrational energy which relaxes into thermal form only during the post-discharge phase. Furthermore, the glow corona discharges that persist during the post-discharge phases induce the so called electrical wind. In the present work, the post-discharge phase is simulated in a bi-dimensional geometry using the commercial FLUENT software. The discharges effects (radical and excited species creation and thermal relaxation) are involved through chemical and energy source terms calculated from a complete 2Drz streamer model already developed.

## 1. INTRODUCTION

Following the increase of atmospheric pollution, mainly due to combustion gases emitted by industry and car engines, several studies focus on pollution control using non-thermal plasma devices. Their advantage is to efficiently remove, several harmful oxides (such as $NO_x$ and $SO_2$). It has been shown that the non-thermal plasma technique is well adapted when oxides are present in very low concentration, like in the case of oil or coal burning flue gases [1]. The gas pollution treatment by corona discharge involves three main steps corresponding to distinct time scales: a) the discharge phase lasts some nanoseconds and corresponds to the formation of primary, by electron-molecule interactions, and secondary radicals, by electron molecule and ion-molecule interactions b) the post-discharge phase where the formed radicals (O, OH, N, $O_3$, $HO_2$) collide with oxides (e.g., NO) to form acids (e.g., $HNO_3$) c) during the last step, these acids interact with a base to form mineral salts and other harmless particles (such as O, N, $N_2$). During the process of pollution control, the time required to remove oxides by reactions with radicals is larger than the radial expansion than the diffusion time of the main chemical species. Therefore, the radial expansion of the gas mixture inside the discharge channel affects significantly the chemical kinetics [2]. The resulting gas temperature rise can largely modify the chemical kinetics and the diffusion phenomena [2,3]. Recently Eichwald and al [4] have looked at the diffusion phenomena and the gas temperature variation associated to the pulsed corona discharge using a 1D model.
The aim of the present work is to couple the chemical and energetic behaviour of the electrical discharge with the airflow at atmospheric pressure through a 2D model of the reactor, allowing to estimate with better precision the areas of formation and diffusion of the new chemical species, including ozone, and to estimate the gas temperature rise during the post-discharge phase. The discharge is simulated by the injection of temporal and local source terms obtained from previous calculations [5,6]. This paper is presented as follows: Section 2 is devoted to the description of the gas dynamics model involving the classical conservation equations and the numerical method used. The source terms, boundary conditions and the selected reactional scheme are provided in Section 3. Section 4 compares the obtained results with those of direct coupling [5,6] and analyses the effect of velocity on the temperature and chemical kinetics spatial distributions.

## 2. POST-DISCHARGE MODEL

The 2D flow model was developed using the commercial *Fluent* software which uses the finite volume method to solve the equations of fluid dynamics. This model takes into account the conservation equations of mass, momentum, and enthalpy, coupled with the vibrational energy density $\varepsilon_v$ conservation (Eq.4):

$$\frac{\partial(\rho u_i)}{\partial t}+\frac{\partial(\rho u_i u_j)}{\partial x_i}=-\frac{\partial p}{\partial x_i}-\frac{\partial \tau_{ij}}{\partial x_j}+\rho g_i \quad (1)$$

$$\frac{\partial(\rho m_{i'})}{\partial t}+\frac{\partial(\rho u_i m_{i'})}{\partial x_i}=\frac{\partial(J_{i',i})}{\partial x_i}+S_{i'} \quad (2)$$

$$\frac{\partial(\rho h)}{\partial t}+\frac{\partial(\rho u_i h)}{\partial x_i}=\frac{\partial}{\partial x_i}\left(k\frac{\partial T}{\partial x_i}\right)-\frac{\partial}{\partial x_i}\sum_{j'=1}^{n}h_{j'}J_{j'}+\frac{\partial p}{\partial t}+ u_i\frac{\partial p}{\partial x_i}+\tau_{ij}\frac{\partial u_i}{\partial x_j}+S_h \quad (3)$$

The vibrational relaxation is described by,

$$\frac{\partial \varepsilon_v}{\partial t}=f_v\mathbf{J.E}-\frac{\varepsilon_v}{\tau_v} \quad (4)$$

$\rho$ mass density of gas, $u_i$ velocity in the i direction, p static pressure, $m_{i'}$ mass fraction of species $i'$, $\tau_{ij}$ is the stress tensor, $g_i$ is the gravitational acceleration, $J_{i',i}$ is the diffusive mass flux of species $i'$ in the i$^{th}$ direction, $S_{i'}$ is the net rate of production of species $i'$ per unit volume (due to chemical reactions and injection of radicals in the discharge zone), $h$ is the static enthalpy, $T$ is the temperature, $J_{j'}$ is the flux of species $j'$, $S_h=(f_{ex}+f_t)\mathbf{J.E}+\frac{\varepsilon_v}{\tau_v}$ is the enthalpy source term, $J$ the current density, $E$ the electric field and $k$ is the gas thermal conductivity. $f_t$ is the fraction of dissipated power $J.E$ used for the direct heating via elastic and rotation collisions. $f_{ex}$ corresponds to the energy lost during the excitation process of the electronic states. This energy is assumed to relax instantaneously into a thermal from. $f_v$ is the fraction of the energy lost during the vibrational excitation and $\tau_v$ is the vibration relaxation time ($\tau_v$=50µs). The different source terms ($S_{i'}$, $S_h$ and vibrational relaxation term) are derived from the model of the corona discharge developed by Ducasse and al [5,6] using the direct coupling. They are averaged over the discharge diameter (50µm) and duration (150ns). Transport coefficient were all calculated via kinetic theory and classical mixing lows with the Leonard-Jones parameters taken from table1 [3].

## 3. SIMULATION CONDITIONS

Two flow velocities (u = 0.5m / s and u = 5m / s) of synthetic air (80% N2 and 20% O2) were simulated. The discharge corresponds to a point-plane electrode configuration with an interelectrode distance of 7mm and a point radius of 25µm. The computational cell is 1cm high and 2.5mm wide, it is meshed with a 200x50 grid. Source terms simulating the discharge are injected at the center of the computational cell in a column of 140 cells, each of them being 50µm wide and high. The discharge phase duration is fixed at 150ns, it is assumed to inject primary radicals (N end O) and enthalpy source terms.

The post-discharge reactive gas dynamics model takes into account 7 neutral chemical species (N, O, O3, NO2, NO, O2 et N2) reacting following 9 selected chemical reactions (table 1).

| Reactions | $K_i$ (cm$^3$ s$^{-1}$, cm$^6$ s$^{-1}$) | $\theta_i$ | $\eta_i$ |
|---|---|---|---|
| $N + O_2 \rightarrow NO + O$ | 0.440 10$^{-11}$ | 0.322 10$^4$ | 0 |
| $N + NO \rightarrow N_2 + O$ | 0.830 10$^{10}$ | 0 | 0 |
| $O + N + N_2 \rightarrow NO + N_2$ | 0.180 10$^{-30}$ | 0 | - 0.5 |
| $O + NO + N_2 \rightarrow NO_2 + N_2$ | 0.175 10$^{-27}$ | 0 | - 1.37 |
| $O_3 + NO \rightarrow O_2 + NO_2$ | 0.18 10$^{-11}$ | 1.137 10$^4$ | 0 |
| $O + NO_2 \rightarrow NO + O_2$ | 0.521 10$^{-11}$ | -0.202 10$^3$ | 0 |
| $N + N + N_2 \rightarrow N_2 + N_2$ | 0.830 10$^{-33}$ | -0.500 10$^3$ | 0 |
| $O + O_2 + N_2 \rightarrow O_3 + N_2$ | 0.300 10$^{-27}$ | 0 | - 2.3 |
| $O + O_2 + O_2 \rightarrow O_3 + O_2$ | 0.300 10$^{-27}$ | 0 | - 2.3 |

*Table 1: Reactions considered in the post-discharge model with parameters of the rate coefficient $R_i(T)$ [3] given following an Arrhenius form: $R_i(T) = K_i T^{\eta_i} e^{-\theta_i/T}$.*

## 4. RESULTS AND DISCUSSION

*A. Discharge phase results (0 < t <150 ns)*

*4.1 Comparison between the corona discharge model and the Fluent simulation*

The comparison at the end of the discharge phase shows that the spatial distributions of *T* and of the atomic species densities are in good agreement (Fig.1).

Figure 2 describes the evolution of temperature and radical densities in the computational cell at the end of the discharge phase. Since the duration of this phase is very short, there is no effect of the flow velocity u on the spatial distributions. The temperature is very high at the electrode point (1102 K) and decreases gradually along the axis. We also note that the densities of O and N follow the same behaviour as T and reached at the point 1.41 $10^{23}$ $m^{-3}$ and 3.02 $10^{22}$ $m^{-3}$ respectively.

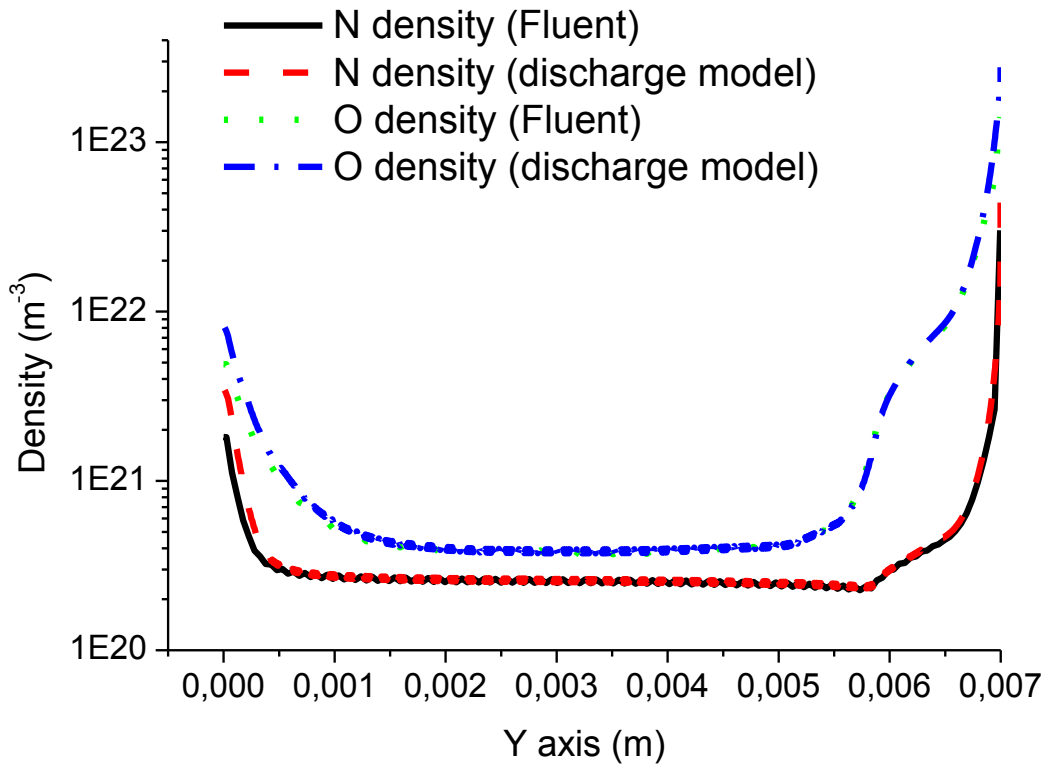

Fig.1: Comparison of the atomic species densities, on the discharge axis, given by the discharge model [5,6] and Fluent after 150ns.

*B. Post-discharge results (150 ns < t < 1 ms)*

*4.2 Temperature relaxation*

Figures 3 (a) and 4 (a) show the evolution of the temperature in the computational cell at 30µs and 1ms. At 30µs, when the airflow velocity is low (0.5m/s), the flow behavior is only governed by diffusion while transport effects begin to appear at higher velocity (5m/s). The temperature strongly decreases on the discharge axis and drops below 380 K close to the anode. After 1ms, the gas temperature in the cell is almost returned to its initial configuration in both cases.

The gas temperature rise due to the vibrational relaxation is shifted in time ($\tau_v$=50µs) and is very small (less than 10K).

4.3 Chemical kinetics analysis

Figures 3 and 4 describe the evolution of the radical species densities of O, N and $O_3$ in the computational cell at 30µs and 1ms. At 30µs, the chemistry is also governed by diffusion with a small transport effect if u=5m/s. The remaining O-atom concentration represents approximately 1% of this initial value. This consummation is due mainly to the formation of ozone by two reactions (8) and (9), whose density is maximum near the anode (2.11 $10^{22}$ m-3) (Figure 3 (d)). The oxygen atoms are also consumed by reaction (3) to form NO, that is oxidised by $O_3$ to form $NO_2$ following reaction 5 (see figure 5).

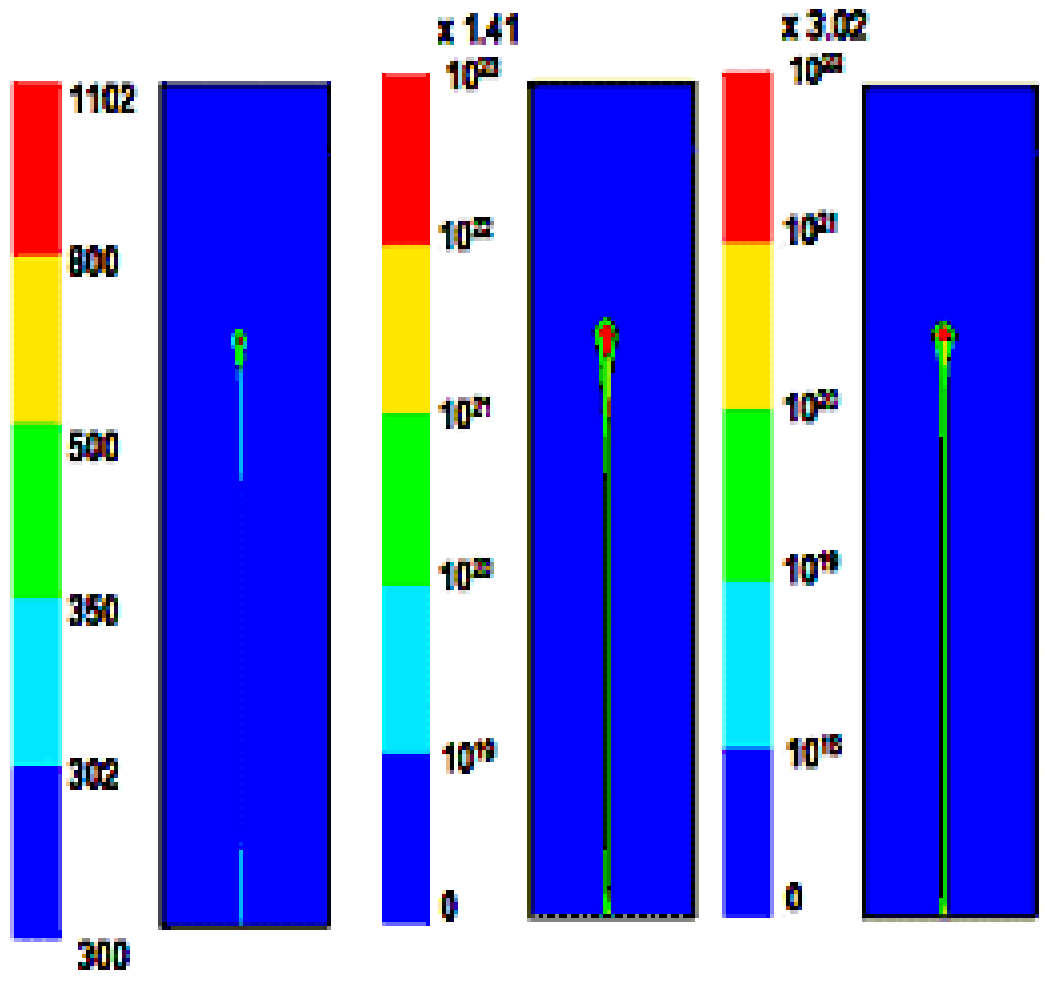

(a) (b) (c)

*Fig.2 : Evolution of (a) temperature,(b) O density ($m^{-3}$). (c) N density ($m^{-3}$). in the computational cell at 150ns. Distributions are identical for u=0.5m/s and u=5m/s.*

## 5. CONCLUSION

This work has demonstrated the feasibility of a coupling between the hydrodynamic of a pollution control reactor and mono filamentary streamer. This coupling can be taken into account satisfactorily by a commercial CFD software on time scales appropriated to the treatment of streamers. The agreement between the obtained results and the one given by a more complete model is good. It is also demonstrated that that primary and secondary radicals can diffuse in volumes...greeter than the discharge one.

This work will be pursued in real 3D conditions by integrating multiple discharge and post-discharge cycles and for different multiple geometries.

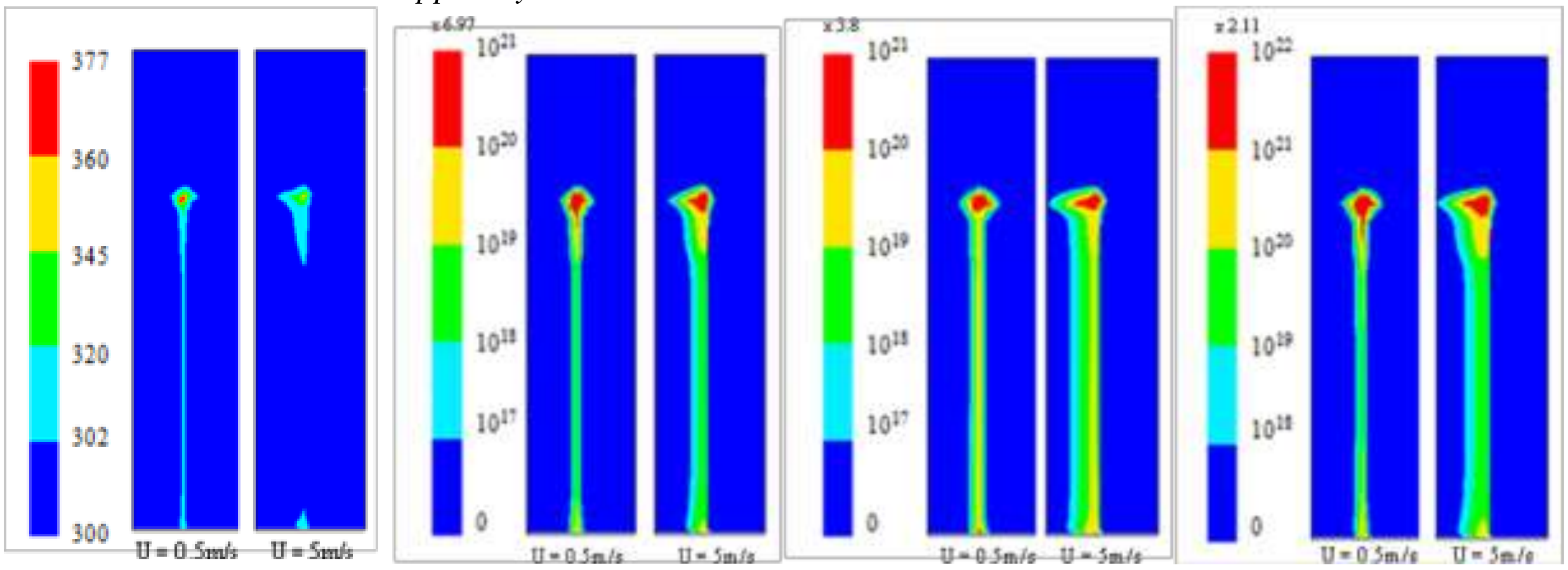

(a) (b) (c) (d)

*Fig.3 : Evolution of (a) temperature (K), (b) O density ($m^{-3}$), (c) N density ($m^{-3}$) and (d) $O_3$ density ($m^{-3}$) densities at t=30µs.*

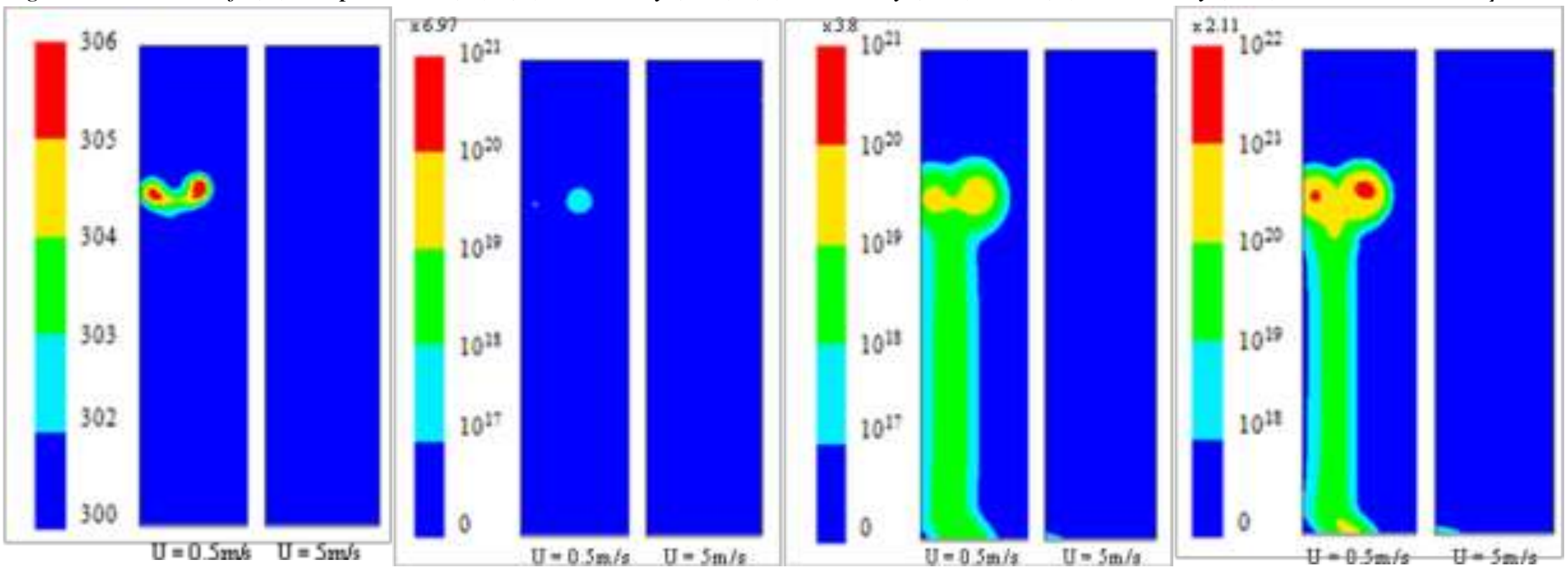

(a) (b) (c) (d)

*Fig.4 : Evolution of (a) temperature (K), (b) O density ($m^{-3}$), (c) N density ($m^{-3}$) and (d) $O_3$density ($m^{-3}$) at t=1ms.*

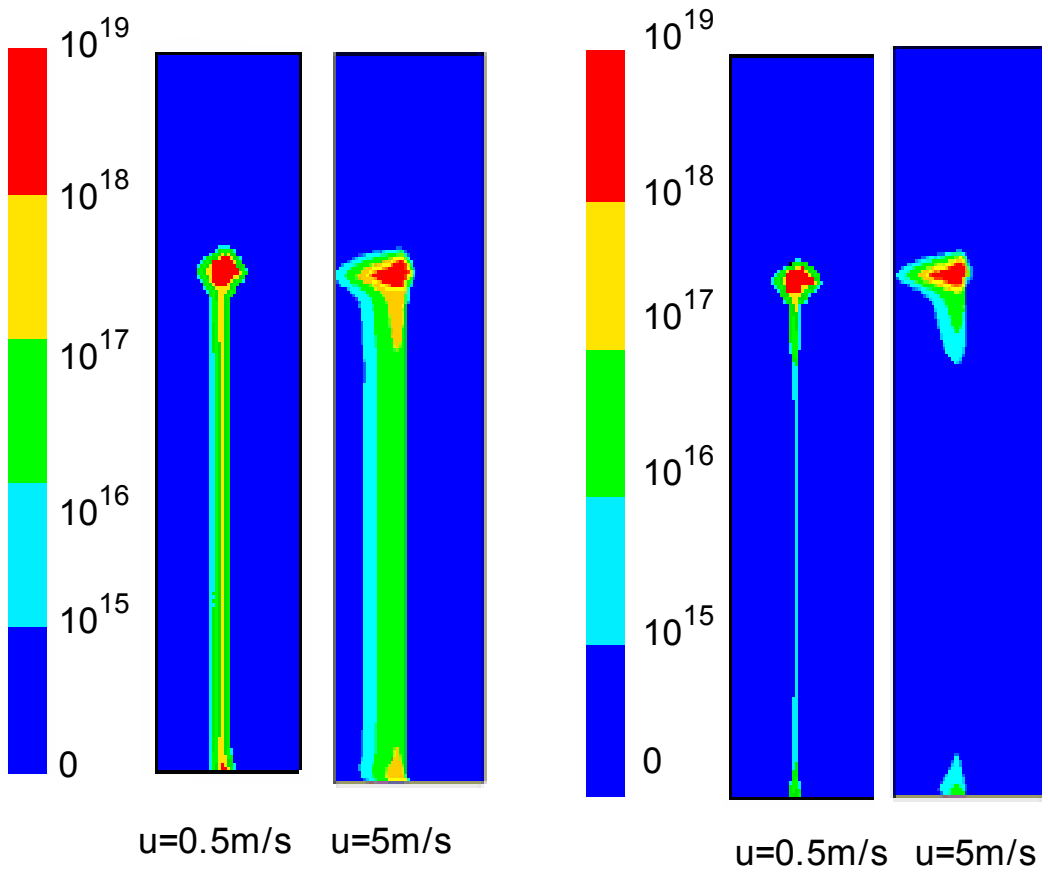

(a) (b)

*Fig.5 : Evolution of (a) NO density ($m^{-3}$), (c) $NO_2$ density ($m^{-3}$) at t=30µs*